\documentclass[useAMS,usenatbib,referee]{biom_altered}
\usepackage{epsfig,rotating}
\usepackage{psfrag} 
\usepackage{subfigure}
\usepackage{graphicx}
\usepackage{natbib} 
\usepackage{amsfonts}
\usepackage{amsmath}
\usepackage{amssymb}
\usepackage[dvips, colorlinks=true]{hyperref}

\usepackage{color}

\newcommand{\Expect}[1]{\mathbb{E} \left[{#1}\right]}

\newcommand{\Var}[1]{\mbox{Var} \left[{#1}\right]}

\newcommand{\Norm}[1]{\left\vert\left\vert{#1}\right\vert\right\vert}

\newcommand{\bma}{\mathbf{a}}
\newcommand{\bmA}{\mathbf{A}}

\newcommand{\bmc}{\mathbf{c}}

\newcommand{\bmF}{\mathbf{F}}

\newcommand{\bmh}{\mathbf{h}}
\newcommand{\bmH}{\mathbf{H}}

\newcommand{\bmm}{\mathbf{m}}
\newcommand{\bmM}{\mathbf{M}}

\newcommand{\bmP}{\mathbf{P}}

\newcommand{\bmS}{\mathbf{S}}

\newcommand{\bmV}{\mathbf{V}}

\newcommand{\bmW}{\mathbf{W}}
\newcommand{\bmx}{\mathbf{x}}
\newcommand{\bmX}{\mathbf{X}}
\newcommand{\bmy}{\mathbf{y}}
\newcommand{\bmY}{\mathbf{Y}}

\newcommand{\bmzero}{\mathbf{0}}

\newcommand{\bmeta}{\mbox{\boldmath$\eta$}}

\newcommand{\bmmu}{\mbox{\boldmath$\mu$}}

\newcommand{\bmPsi}{\mbox{\boldmath$\Psi$}}

\newcommand{\bmtheta}{\mbox{\boldmath$\theta$}}

\newcommand{\bmSigma}{\mbox{\boldmath$\Sigma$}}

\newcommand{\oft}{(t)}

\newcommand{\oftime}[1]{({#1})}

\newcommand{\sPred}{\mbox{Pred}}
\newcommand{\sPrey}{\mbox{Prey}}
\newcommand{\sempty}{\phi}

\newcommand{\sDNA}{\mbox{DNA}}
\newcommand{\sP}{\mbox{P}}
\newcommand{\sPP}{\mbox{P}_2}
\newcommand{\sRNA}{\mbox{RNA}}

\title[Inference using the LNA]{Inference for reaction networks using the
  Linear Noise Approximation}

\author{
Paul Fearnhead$^{1,*}$
\email{p.fearnhead@lancaster.ac.uk},
Vasilieos Giagos$^{2,**}$
\email{v.giagos@kent.ac.uk}
, and
Chris Sherlock$^{1,***}$\email{c.sherlock@lancaster.ac.uk}\\
$^{1}$Department of Mathematics and Statistics, Lancaster University,
Lancaster, LA1 4YF, UK.\\
$^{2}$School of Mathematics, Statistics and Actuarial Science, University
of Kent, Canterbury, Kent CT2 7NF, UK.}

\begin{document}
\date{SomeDate}

\pagerange{\pageref{firstpage}--\pageref{lastpage}} 
\volume{TBA}
\pubyear{2014}
\artmonth{TBA}
\doi{10.1111/biom.12152}

\label{firstpage}

\begin{abstract}
We consider inference for the reaction rates in discretely
observed networks such as those found in models for systems biology,
population ecology and epidemics. Most such networks are neither slow
enough nor small enough for inference via the true state-dependent
Markov jump process to be feasible.  Typically, inference is
conducted by approximating the dynamics through an ordinary differential equation (ODE), or a stochastic differential equation (SDE). The former ignores the stochasticity in the true model, and can lead to inaccurate inferences. The latter is more accurate but is harder to implement as the transition density of the SDE model is generally unknown. The Linear Noise Approximation
(LNA) arises from a first order Taylor expansion of the approximating SDE about
a deterministic solution and can be viewed as a compromise between the ODE and SDE models. It is a stochastic model, but discrete time transition probabilities for
the LNA are available through the solution of a series of ordinary
differential equations. We describe how a restarting LNA can be efficiently 
used to perform inference for a general class of reaction networks; evaluate the accuracy of such an approach; and show how and when this approach is 
either statistically or computationally more efficient than ODE or SDE
methods. We apply the LNA to analyse Google Flu Trends data from the North and South Islands of New Zealand, and are able to
obtain more accurate short-term forecasts of new flu cases than
another recently proposed method, although at a greater computational cost.
\end{abstract}


\maketitle

\section{Introduction}
Reaction networks are used to model a wide variety of real-world
phenomena; they describe a probabilistic mechanism for the joint evolution of
one or more populations of \textit{species}. These \textit{species} may be biological
species,  such as a range of different proteins \cite[e.g.][]{GolightlyWilkinson:2005,GolightlyWilkinson:2008,Boysetal:2008,Fermetal:2008,Proctoretal:2005},
 animal species, such
as predators and their prey \cite[e.g][]{Boysetal:2008,Fermetal:2008}, or
interacting groups of humans or animals such as those infected with a
particular disease, those susceptible to the disease and those who
have recovered from it \cite[e.g][]{AnderssonBritton:2000,BallNeal:2008,JewellKeelingRoberts:2009}.   

The evolution of these networks is most naturally modelled via a continuous-time
jump Markov process. The current \textit{state} of the system is encapsulated in a vector
giving the numbers of each species that are present. The evolution
of the state is described by a series of {\em reactions}, such as the interaction of two
copies of a protein producing a dimer of that protein; or an interaction between an
infected individual and a susceptible individual resulting in the susceptible becoming infected.
Occurences of a given reaction are modelled as a
Poisson process, the rate of which depends on the current state of the
system. Interest lies in inferring the parameters that govern the rate of each
reaction from data on the evolution of the system.

This article concerns inference for the rate parameters and prediction
of the future state in discretely
observed reaction networks allowing that observations may contain
noise and may be of only a subset of the species in the system. 
Inference under the jump Markov process is possible only for
networks which involve few species, few reactions, and
not ``too many'' transitions between observations
\cite[e.g.][]{Boysetal:2008,Amrein:2012}. For most systems it is necessary
to approximate the evolution of the process to make  inference 
computationally feasible. Often this will involve approximating
the evolution through a system of ordinary differential equations 
\cite[ODEs, e.g.][]{JonesPlankSleeman:2010} or stochastic differential equations \cite[SDEs, e.g.][]{Wilkinson:2006}. Models
based on ODEs are only appropriate for very large systems for which the
stochasticity in the evolution is small. For medium-size systems
\cite{Wilkinson:2006} shows that SDE models are more appropriate and can lead to
sensible estimates of the reaction rates. However inference for SDE models
is non-trivial, as the transition density of general SDEs is unknown.

In this paper we use an alternative approximation, known as the Linear Noise Approximation (LNA) 
\cite[]{vanKampen:1997,Elf:2003,Hayot:2004,Fermetal:2008}.
The LNA is obtained through first approximating the dynamics by a system of ODEs, and then modelling
the evolution of the state about the deterministic solution of the ODE, through a linear SDE. Simulations
suggest that this approach has similar accuracy to modelling the system directly through SDEs, but it has the
important advantage that under the LNA the stochastic model for the
states is a Gaussian process. Transition densities are therefore Gaussian, and their mean and covariance can be obtained through solving a system
of differential equations. Using the LNA can therefore be more accurate than modelling the system via ODEs, 
and it is substantially easier to perform inference than under a general SDE model.

The structure of the paper is as follows. 
The next section provides more information on reaction
networks and details two particular examples that will be revisited:
the Lotka-Volterra predator-prey system, and an autoregulatory gene network. 
Section \ref{sect.approx} examines the different possible approximations to the
evolution of reaction networks: ODE approximation; SDE approximation
and the LNA. 
Section
\ref{sect.lnakf} shows how we can calculate likelihoods using a `restarting' LNA for a range of observation models, 
and suggests a simple way of embedding this within MCMC to perform
Bayesian inference.
We evaluate the use of the LNA empirically on both simulated and real data.
An alternative use of a (non-restarting) LNA for inference on reaction networks has previously
been suggested by  \cite{Komorowskietal:2009}. 
In Section \ref{sect.simstudy} we compare our approach with that of \cite{Komorowskietal:2009}, with
the simple ODE approximation, and 
with the SDE-based algorithm of \cite{GolightlyWilkinson:2005}.
In Section \ref{sec.GFT} we use the LNA to analyse Google Flu Trends data
from New Zealand, comparing the accuracy of week-ahead predictions with those from
the recent approach of \cite{JASA:2012}. 

\section{Reaction networks}
\label{sect.reaction.networks}

Consider a general reaction of the form $B+C\rightarrow D$, where the
number of elements of species $B$ and $C$ are respectively $X_B$ and $X_C$
and where the elements are distributed uniformly at random throughout
some volume of space. The
reaction occurs with some fixed probability whenever an element of
species $B$ is within some ``reaction distance'' of an element of $C$; occurences of the reaction may therefore be
modelled as a Poisson process. With further, system dependent,
assumptions, 
the rate, $h$, of
the process is proportional to $X_BX_C$. 
Applying a similar argument, the rate of a reaction such as $B\rightarrow C$ is simply
proportional to $X_B$ and the rate of $2B\rightarrow C$ is
proportional to $0.5 X_B(X_B-1)$. For a fuller discussion of
mass-action kinetics see, for example, \cite{Gillespie:1992}.

Consider now a network of $n_r$ such reactions each involving at least one of 
the $n_s$ species in the population. The dynamics of this model can be
described by a vector of rates of the reactions together with a matrix which describes the effect of
each reaction on the state. We denote by $\bmh$ the $n_r$-vector of reaction
rates. Now define $A_{ij}$ be the net effect on species $j$ of a
single occurence of reaction $i$: so $A_{ij}=0$ means that the number of species $j$
is unaffected by reaction $i$, whereas $A_{ij}=1$ (or $-1$) means the number of species $j$ will 
increase (or decrease) by 1.  The $n_r\times n_s$ 
matrix $\bmA$ is known as the net effect matrix. An equivalent way of defining the effect of 
a set of reactions is via the stoichiometry matrix, $\bmA'$, where
throughout this article, $'$ denotes the transpose of a matrix.

{\bf Example 1: the Lotka-Volterra model}

The Lotka-Volterra model 
\cite[e.g.][]{Wilkinson:2006} describes a population of
two competing species: \textit{predators} which die with rate $\theta_2$
and reproduce with rate $\theta_1$ by consuming \textit{prey} which
reproduce with rate $\theta_3$. In its simplest form the probabilistic
system is defined by:
\[
R_1:\sPred+\sPrey\rightarrow 2\sPred;~~
R_2:\sPred\rightarrow \sempty;~~
R_3:\sPrey\rightarrow 2\sPrey.
\]
Denoting the number of $\sPred$ by $X_1$ and the number of
$\sPrey$ by $X_2$ gives the vector of
reactions rates and the net effect matrix, respectively as:
\renewcommand{\arraystretch}{0.8}
\[
\bmh:=\left(\theta_1 X_1 X_2,~\theta_2 X_1,~\theta_3 X_2\right)
~~~\mbox{and}~~~
\bmA'=
\left[
\begin{array}{rrr}
1&-1&0\\
-1&0&1
\end{array}
\right].
\]  

{\bf Example 2: autoregulatory gene network}

The following system describes the self-regulating production of a protein, $\sP$, and its dimer,
$\sPP$. The system is analysed in
\cite{GolightlyWilkinson:2005} and is also discussed in
\cite{Wilkinson:2006}, while a similar system is analysed in
\cite{GolightlyWilkinson:2008}. 
Reactions $R_1$ and $R_2$ describe the reversible process
whereby the protein dimer $\sPP$ binds to the gene (which we denote as
$\sDNA$) and thereby inhibits
the production,  by reactions $R_3$ and
$R_4$, of the protein, $\sP$. Dimerisation of the protein and the reverse reaction are
described by Reactions $R_5$ and $R_6$, while $R_7$ and $R_8$ describe
the destruction of the protein and of the enzyme RNA-polymerase, which
is denoted $\sRNA$.
\begin{center}
\begin{tabular}{ll}
$R_1:$ $\sDNA+\sPP\rightarrow \sDNA\cdot \sPP$
&
$R_2:$ $\sDNA\cdot \sPP\rightarrow\sDNA+\sPP$.\\
$R_3:$ $\sDNA\rightarrow \sDNA+\sRNA$&
$R_4:$ $\sRNA\rightarrow \sRNA+\sP$.\\
$R_5:$ $2\sP\rightarrow \sPP$&
$R_6:$ $\sPP \rightarrow 2\sP$\\
$R_7:$ $\sRNA\rightarrow 0$&
$R_8:$ $\sP\rightarrow 0$.
\end{tabular}
\end{center}

\vspace{.5cm}

From the reactions, the total, $k$, of the number of $\sDNA$ and $\sDNA\cdot\sP2$ molecules
is fixed throughout the evolution of the system. Denoting the number of molecules of $\sDNA$, $\sRNA$, $\sP$, and
$\sP_2$ as $X_1,~X_2, ~X_3,$ and $X_4$ respectively therefore leads to
a reaction rate vector of $\bmh:=(\theta_1 X_1 X_4,~\theta_2 (k-X_1),~\theta_3 X_1, ~\theta_4 X_2, ~\theta_5X_3(X_3-1)/2,~\theta_6X_4,~\theta_7X_2,~\theta_8X_3)$.
The net effect matrix for this example is $\bmA$, where
\[
\bmA'=
\left[
\begin{array}{rrrrrrrr}
-1&1&0&0&0&0&0&0\\
0&0&1&0&0&0&-1&0\\
0&0&0&1&-2&2&0&-1\\
-1&1&0&0&1&-1&0&0
\end{array}
\right].
\]
Further examples, of a one and two-island epidemic model, are detailed in
  Appendix A.
 
\section{Approximations for network evolution}
\label{sect.approx}
We first consider the ODE and SDE approximations to the true
process, and then
sketch the justification for the Linear Noise Approximation (LNA).

It will be helpful to denote the $n_s$-vector holding the number of molecules of each
species by $\bmX$ and to 
define the $n_r\times n_r$ reaction rate matrix $\bmH:=\mbox{diag}(\bmh)$.

\subsection{The ODE and SDE approximations}
\label{sect.cla}

In an infinitesimal time $dt$ the mean and variance of the change in
$\bmX$ due to all of the $n_r$ independent Poisson processes can be calculated as \cite[e.g.][]{Wilkinson:2006}:
\[
\Expect{d\bmX\oft}=\bmA'\bmh ~dt,
~~~
\Var{d\bmX\oft}=\bmA' \bmH \bmA ~dt.
\]

The ODE approximation to the evolution ignores the stochasticity of the model and is based solely on
the expected change in the mean. This gives the following differential equation
\[
\frac{\mbox{d}\bmX\oft}{\mbox{d}t}=\bmA'\bmh(\bmX\oft,\bmtheta).
\]
The SDE approximation models stochasticity through 
\[
d\bmX\oft=\bmA'\bmh(\bmX\oft,\bmtheta)dt +
\sqrt{\bmA'\bmH(\bmX\oft,\bmtheta)\bmA} ~d\bmW\oft,
\]
where the matrix $\sqrt{\bmA'\bmH(\bmX\oft,\bmc)\bmA}$ 
is any (without loss of generality, $n_s\times n_s$) matrix square root,
such as that obtained by Cholesky 
decomposition, and $\bmW\oft$ is
Brownian motion.

The ODE model is deterministic, and fitting the model generally involves estimating both the
initial condition and parameter values that give the best fit to the data. Often the fit to the data
is quantified by the sum of the square residuals \cite[see][and references therein]{Ramsay:2007}. 

There are a range of methods for estimating parameters of an SDE model \cite[see e.g.][]{Sorensen:2004}.
Recently, there has been much research on how to implement likelihood-based methods 
\cite[e.g.][]{Elerian:2001,Durham/Gallant:2002,Beskos/Papaspiliopoulos/Roberts/Fearnhead:2006,Ait-Sahalia:2008}.
Generally, however, the SDE model will not lead to a tractable 
distribution for $\bmX\oft$ given $\bmX_0$, and hence these models
have an intractable likelihood. To overcome this complication it
is common to approximate the transition density of the SDE, for example by the Euler approximation \cite[]{Kloeden/Platen:1992}.
The Euler approximation is only accurate over small time-intervals. The implementation of  these methods therefore
involves discretising time between each observation, and using computationally-intensive methods that 
impute values of the state at both the observation times and the grid of times between each observation. 
For example  \cite{GolightlyWilkinson:2005} implement such a method with an MCMC scheme, and \cite{GolightlyWilkinson:2006b} within
a sequential Monte Carlo algorithm. There is a considerable computational overhead in implementing these methods which
increases with the fineness of the grid of time-points between observations, and this  has led to much research on efficient MCMC and other methods. See \cite{RobertsStramer:2001} for a discussion
of how the fineness of the grid can affect mixing of the MCMC and, for example, \cite{GolightlyWilkinson:2008} for details of more efficient 
MCMC approaches.

\subsection{The Linear Noise Approximation}
\label{sect.intro.thealgo}
The Linear Noise Approximation (LNA) first appeared as a functional central limit law for density
dependent processes; see \cite{Kurtz:1970} and
\cite{Kurtz:1971} for the technical conditions. 
It approximates the dynamics of the network  by an SDE
which has 
 tractable transition densities between observation times; inference therefore
does not require any data augmentation \cite[]{vanKampen:1997}.

Whilst \cite{Kurtz:1970} and \cite{Kurtz:1971} justify our use of the
LNA it will be more helpful in the present context to consider the LNA as a general approximation to the solution
to an SDE, and then apply this to the SDE model derived in the previous section.
The idea of the LNA is that we partition $\bmX\oft$ into a deterministic path,
$\bmeta\oft$, and a stochastic perturbation from this path. Under the
assumption that the perturbation is ``small'' relative to the
deterministic path the
distribution of an approximate solution at any given time
point is found by solving a series of ODEs. In our applications the deterministic path
is just the solution of the ODE model introduced in the previous section.
Here we provide a
 short heuristic motivation of the approximation; for a more rigorous
 derivation and more detailed discussion the reader is referred to
 \cite{Fermetal:2008}.

Consider the general SDE for vector $\bmX$ of length $n_s$
\begin{equation}
\label{eqn.full.sde}
d\bmX\oft = \bma(\bmX\oft) ~dt + \epsilon\bmS(\bmX\oft)~d\bmW\oft,
\end{equation}
with initial condition $\bmX\oftime{0}=\bmX_0$. 
Let $\bmeta\oft$ be the (deterministic) solution to 
\begin{equation}
\label{eqn.deterministic.y}
\frac{d\bmeta}{dt}=\bma(\bmeta)
\end{equation}
with initial value $\bmeta_0$. We assume that over the time interval
of interest $\Norm{\bmX-\bmeta}$ is
$O(\epsilon)$. Set
$\bmM\oft=(\bmX\oft-\bmeta\oft)/\epsilon$ and use a Taylor expansion of $\bma$ and $\bmS$ about $\bmeta\oft$ in (\ref{eqn.full.sde}).
Collecting terms of $O(\epsilon)$ gives
\begin{equation}
\label{eqn.perturb}
d\bmM\oft=\bmF\oft \bmM\oft~dt + \bmS\oft ~d\bmW\oft,
\end{equation}
where $\bmF$ is the $n_s \times n_s$ matrix with components
\[
F_{ij}\oft=\left.\frac{\partial a_i}{\partial x_j}\right|_{\bmeta\oft}
,~~~\mbox{and}~~~
\bmS\oft=\bmS(\bmeta\oft).
\]
The use of $\epsilon$ in (\ref{eqn.full.sde}) is purely to indicate
that the stochastic term $\epsilon\bmS(\bmX\oft)$ is ``small'' relative to the drift, and to aid
in the collection of terms of similar size. Henceforth it will be simpler to set
$\epsilon=1$ and assume that $\bmS(\bmX\oft)$ is ``small''.
The initial condition for (\ref{eqn.perturb}) is therefore
$\bmM_0=(\bmX_0-\bmeta_0)$.  

Provided that $\bmX_0$ has either a point mass at $\bmx_0$ or has a
 Gaussian distribution, the increment in (\ref{eqn.perturb}) is
 a linear combination of Gaussians so $\bmM\oft$ has a Gaussian
 distribution for all $t$.
The mean and variance of this Gaussian can be obtained by
solving a series of ODEs,
\begin{eqnarray}
\label{eqn.lna.mean}
\frac{d\bmm}{dt}&=&\bmF\bmm,\\
\label{eqn.ode.var}
\frac{d\bmPsi}{dt}&=&\bmPsi\bmF^t+\bmF\bmPsi+\bmS\bmS^t,
\end{eqnarray}
where
$\bmm\oft:=\Expect{\bmM\oft}$, and 
$\bmPsi\oft:=\Var{\bmM\oft}$ (see Appendix B for the derivation).

Suppose $\bmX_0\sim N(\bmmu^*_0,\bmSigma^*_0)$, then for
arbitrary $\bmeta_0$ we may set $\bmeta(0)=\bmeta_0$,
$\bmm(0)=\bmmu^*_0-\bmeta_0$ and $\bmPsi(0)=\bmSigma^*_0$. Integrating
(\ref{eqn.deterministic.y}), (\ref{eqn.lna.mean}) and
(\ref{eqn.ode.var}) through to time $t$ provides the LNA approximation
\begin{equation}
\label{eqn.transitionGauss}
\bmX_t\sim N(\bmeta(t)+\bmm(t),~\bmPsi(t)).
\end{equation}
Transition probabilities for the autoregulatory model
(Example 2) given by the LNA and estimated from the SDE approximation
were compared with estimates of the true probabilities for three
different system sizes (see Appendix C for details). The results
suggest that even for relatively small system sizes LNA transition
probabilities are
comparable with those from the SDE and can provide a reasonable
approximation to the probabilities under the MJP.

\section{Inference using the LNA}
\label{sect.lnakf}
We first
briefly describe the inference methodolody when we observe a system
exactly and completely at a discrete set of times. We then
show how to perform inference when only
linear combinations of a subset of species 
are observed, and these, potentially, are observed with error. Finally
we compare 
our approach with an alternative method of using the LNA for inference introduced by \cite{Komorowskietal:2009}.

\subsection{The fully and exactly observed system}
\label{sect.inffullexact}
Consider the situation where at each of a discrete set of times,
$t_i~(i=0,\dots,n)$, the system, $\bmx_i$, is observed completely and
without error. Let the true transition density of the system be denoted by
$\pi(\bmx_i|\bmx_{i-1},\theta)$ and the LNA approximation of this by
$\hat\pi_{LNA}(\bmx_i|\bmx_{i-1},\theta)$. In practice we use the LNA approximation
obtained using $\bmeta=\bmx_i,~\bmm=\bmzero,~\bmPsi=\bmzero$.  This implementation is based on an ODE solution that is piecewise continuous, with discontinuities 
at observation times as we restart each ODE solution at the
observations. Further, as $\bmm(t_i)=\bmzero$, directly from
(\ref{eqn.lna.mean}) we have $\bmm(t)=\bmzero$ for all $t>t_i$.

For a fully observed system, the likelihood factorises as
$ L(\theta)=\prod_{i=1}^n \pi(\bmx_i|\bmx_{i-1},\theta)$.
This motivates using the approximation:
 $\hat{L}_{LNA}(\theta)=\prod_{i=1}^n\hat\pi_{LNA}(\bmx_i|\bmx_{i-1},\theta)$.
The approximation is a product of Normal densities, with the mean and covariances
depending on the parameters. It is possible to maximise this likelihood numerically. Alternatively, if we
introduce priors for $\theta$, $\pi(\theta)$, we can use standard MCMC algorithms to sample from the
corresponding approximation to the posterior which is proportional to $\pi(\theta) \hat{L}_{LNA}(\theta)$.

The accuracy of estimators obtained by maximising $\hat{L}_{LNA}(\theta)$ has been extensively studied in \cite{Giagos:2011}, for
both the Lokta-Volterra model (Example 1), and the auto-regulatory model (Example 2). The method gave reliable point estimates of
parameters, and reasonable estimates of uncertainty (coverage of $95\%$ confidence intervals was generally $90\%$ for small systems,
and close to $95\%$ for large systems).

\subsection{Partially-observed systems}
\label{sect.kalman}
Now assume that we have partial observations $\bmy_0,\dots,\bmy_n$ from
times $0=t_0,\dots,t_n=T$, where the conditional distribution
for the observations given the true process is
\[
\bmY_i|\bmx_i \sim N\left(\bmP(\bmtheta)\bmx_i,\bmV(\bmtheta)\right).
\]
For the examples in Section \ref{sect.simstudy} the matrix $\bmP(\bmtheta)$ simply
removes certain components of $\bmx_i$ and leaves the remaining
components unchanged; an operation that requires no parameterisation,
but for the model analysed in Section \ref{sec.GFT} the 
 observations are centered on an
unknown but fixed multiple of the true values (with
Gaussian error). The variance of the Gaussian error, $\bmV$, can be any
deterministic function (of time, for example) parameterised by
$\bmtheta$. In the examples that we consider in this article
$\bmV$ is either $\bmzero$ or a fixed (unknown) diagonal matrix. We also introduce a prior, 
$\bmX_0\sim N(\bmmu_0,\bmSigma_0)$.
To simplify notation, in the following we drop the explicit dependence of the
matrices $\bmP(\bmtheta)$ and $\bmV(\bmtheta)$ on $\bmtheta$. We will also
use $\bmy_{0:i}:=(\bmy_0,\ldots,\bmy_i)$.

\subsubsection{Approximating the Likelihood using the LNA}
\label{sect.approx.key.steps}
Any likelihood may be decomposed as
\[
 L(\theta)=\pi(\bmy_0|\theta)\prod_{i=1}^n \pi(\bmy_{i}|\bmy_{0:i-1},\theta).
\]
Firstly, $\pi(\bmy_0)$ can be calculated directly from our model as
$\bmY_0 \sim N(\bmP\bmmu_0,\bmP\bmSigma_0\bmP^t+\bmV)$.
We then calculate approximations
$\hat{\pi}_{LNA}(\bmy_{i}|\bmy_{0:i-1},\theta)$ to $\pi(\bmy_{i}|\bmy_{0:i-1},\theta)$
recursively for $i=1,\ldots,n$. 

Standard results give
$\bmX_0|\bmy_0 \sim N(\bmmu^*_0,\bmSigma_0^*),$
where
\begin{eqnarray*}
\bmmu_0^*&=&
\bmmu_0 +
\bmSigma_0\bmP^t\left(\bmP\bmSigma_0\bmP^t+\bmV\right)^{-1}
\left(\bmy_0-\bmP\bmmu_0\right)\\
\bmSigma_0^*&=&
\bmSigma_0-
\bmSigma_0\bmP^t\left(\bmP\bmSigma_0\bmP^t+\bmV\right)^{-1}\bmP\bmSigma_0.
\end{eqnarray*}
We then apply Kalman filter recursions and the LNA, repeating the following steps:
\begin{enumerate}
\item
\textbf{Obtain the predictive distribution at time $t_{i}$}. \\
We will have that for suitable $\bmmu^*_{i-1}$ and $\bmSigma^*_{i-1}$:
\[
  \bmX_{i-1}|\bmy_{0:i-1} \sim N(\bmmu^*_{i-1},\bmSigma^*_{i-1}).
\]
We then initiate the LNA with $\bmeta\oftime{t_{i-1}}=\bmmu_{i-1}^*$, so that 
$\bmm\oftime{t_{i-1}}=\bmzero$, and
$\bmPsi\oftime{t_{i-1}}=\bmSigma^*_{i-1}$.  From (\ref{eqn.lna.mean}),
$\bmm\oftime{t_i}=\bmzero\Rightarrow \bmm\oftime{t}=\bmzero$ for all
$t>t_i$. Further,  
integrating the ODEs (\ref{eqn.deterministic.y}) and (\ref{eqn.ode.var})
forward for time
$t_{i}-t_{i-1}$ provides $\bmeta\oftime{t_{i}}$ and 
$\bmPsi\oftime{t_{i}}$, so that our approximation to the density at
$t_i$ is  
$\bmX_{i}|\bmy_{1:i-1}\sim N\left(\bmmu_{i},\bmSigma_{i}\right)$,
where $\bmmu_{i}=\bmeta\oftime{t_{i}}$ and $\bmSigma_{i}=\bmPsi\oftime{t_{i}}$.
\item
\textbf{Calculate $\hat{\pi}_{LNA}(\bmy_{i}|\bmy_{0:(i-1)},\theta)$}. \\
Using $\bmY_i=\bmP\bmX_i+\epsilon_i$, where $\epsilon_i\sim N(0,\bmV)$ directly gives 
\[
\bmY_{i}|\bmy_{0:i-1}\sim 
N\left(
\bmP\bmmu_{i},
\bmP\bmSigma_i\bmP^t+\bmV
\right).
\]
\item \textbf{Calculate $\hat{\pi}_{LNA}(\bmx_i|\bmy_{0:i},\theta)$.}\\
Since 
\begin{equation}
\left[
\begin{array}{c}
\bmX_{i}\\
\bmY_{i}
\end{array}
\right]
\left| \right.
\bmy_{1:(i-1)}
\sim
N\left(
\left[
\begin{array}{l}
\bmmu_{i}\\
\bmP\bmmu_{i}
\end{array}
\right]
,
\left[
\begin{array}{ll}
\bmSigma_i&\bmSigma_i\bmP^t\\
\bmP\bmSigma_{i}&\bmP\bmSigma_i\bmP^t+\bmV
\end{array}
\right]
\right)
\label{eqn.joint.kalman}
\end{equation}
we have directly that $\bmX_i|\bmy_{1:i}\sim
N(\bmmu_i^*,\bmSigma_i^*)$, where
\begin{eqnarray}
\bmmu_i^*&=&
\bmmu_i +
\bmSigma_i\bmP^t\left(\bmP\bmSigma_i\bmP^t+\bmV\right)^{-1}
\left(\bmy_i-\bmP\bmmu_i\right)\\
\bmSigma_i^*&=&
\bmSigma_i-
\bmSigma_i\bmP^t\left(\bmP\bmSigma_i\bmP^t+\bmV\right)^{-1}\bmP\bmSigma_i.
\end{eqnarray}
\end{enumerate}
Our approximation
for the likelihood of the data is then
\begin{equation} \label{LNAlik}
 \hat{L}_{LNA}(\bmtheta)=\pi(\bmy_0|\theta)\prod_{i=1}^n \hat\pi_{LNA}(\bmy_{i}|\bmy_{0:i-1},\theta).
\end{equation}
The only approximation in $\hat{L}_{LNA}(\bmtheta)$ is due to using the LNA for the transition density of the system, which gives us the Normal approximation to $\bmX_{i}|\bmy_{1:i-1}$ from the Normal approximation to 
$\bmX_{i-1}|\bmy_{1:i-1}$. 

We emphasise that in Step (1), once the observation $\bmy_{i-1}$
is available then for the
  period of integration from $t_{i-1}$ to $t_{i}$ we
  \textit{re-initialise} the ODE (\ref{eqn.deterministic.y}) to the posterior mean
  at $t_{i-1}$ by setting $\eta(t_{i-1})=\mu^*_{i-1}$, leading to a
  piecewise-continuous solution for $\eta$. The LNA relies on a first-order Taylor
expansion about $\eta$, and by continually realigning the point about
which the expansion is performed to the current best estimate of the
centre of the distribution we aim to minimise the impact of the
higher-order
 terms that have been neglected.

\subsection{MCMC scheme}
It is possible to estimate the parameters by numerically maximising the LNA approximation to the likelihood (\ref{LNAlik}). However we consider a Bayesian analysis. 
We introduce priors for the parameters, $\pi(\bmtheta)$, and use MCMC
to generate samples from the resulting approximation to the posterior
$\pi(\bmtheta)\hat{L}_{LNA}(\bmtheta)$.

We implemented a random-walk Metropolis algorithm (RWM). Each
iteration of the algorithm involved a single block update of all the
log-parameters.  Using the log-scale is natural as all parameters are positive.
 For the simulation study in
 Section \ref{sect.simstudy} we used pilot runs to tune our
 algorithms (and algorithms against which we compare): 
the covariance of the random-walk proposal was proportional to the
  estimate of the covariance of the posterior from the pilot run, with the scale tuned to produce an
acceptance rate in the range 0.25--0.30
\cite[]{RobertsRosenthal:2001}. For the analysis of the Google Flu
Trends Data in Section \ref{sec.GFT} we used an adaptive RWM algorithm
similar to that in \cite{Sherlock/Fearnhead/Roberts:2010}.

\subsection{Implementation}

The ODEs required for calculating $\pi(\bmy_{i}|\bmy_{0:(i-1)})$ can be solved numerically. Care is needed as in many applications the ODEs are stiff \cite[]{hairer1991}. 
There are standard numerical routines for solving stiff ODEs, and we
used the \texttt{lsoda} package \cite[]{LSODA}. 

\subsection{Alternative use of the LNA}
\label{sect.lnakom}

Previously, use of the LNA for Bayesian inference on stochastic kinetic
networks has been suggested 
by  \cite{Komorowskietal:2009}, but their implementation has important differences from ours. The approach
of \cite{Komorowskietal:2009} involves using the LNA to obtain an approximation for the joint distribution of
$\bmX_{1:n}=(\bmX_1,\ldots,\bmX_n)$ conditional on a value for $\bmx_0$. This can be combined with the linear-Gaussian
relationship between each observation $\bmY_i$ and state-value $\bmX_i$ to give an approximation to the
 likelihood for data $\bmy_{1:n}$ in terms of the parameters, $\bmtheta$ and the initial value, $\bmx_0$. They introduce priors
for the $\bmtheta$ and $\bmx_0$, and sample from the (approximate) posterior for these using MCMC.

In practice the most important difference between this approach and ours, is that \cite{Komorowskietal:2009} use the LNA over a time period $[0,t_n]$ obtained
from solving the ODE approximation to the model over this period for a given initial condition. By comparison we use a different LNA
for each time-interval $[t_{i-1},t_i]$, essentially restarting the LNA
using the posterior mean of $\bmx_{t-1}$ given $\bmy_{0:t-1}$
as the initial condition for the ODE (\ref{eqn.deterministic.y}). This difference can be important for some models, as the ODE solution can become poor over long
time-periods. Thus the approach of \cite{Komorowskietal:2009} can give a poor approximation to the distribution of $\bmX_t$ for larger 
values of $t$. By re-starting the LNA method over each time-interval we help avoid the problems of the approximation getting worse
for larger $t$. 

The difference in accuracy of the two approaches for using the LNA is investigated thoroughly for systems which are fully observed 
at discrete time-points in \cite{Giagos:2011}, where the method of
\cite{Komorowskietal:2009} was found to be less accurate, for both point and interval estimation, than the method we
introduce above. We further demonstrate the increased accuracy of our approach for partially-observed systems in Section \ref{sect.simstudy}.

\section{Simulation study}
\label{sect.simstudy}

We now empirically evaluate the performance of the LNA for inference on parameters in both the 
Lotka-Volterra model (Example 1) and the auto-regulatory model (Example 2). Our aim is to both compare our approach 
with inference based on ODE approximations, the LNA approach of \cite{Komorowskietal:2009}, and the SDE-based 
approach of \cite{GolightlyWilkinson:2005}; and to evaluate the
accuracy obtained by using the LNA for both point and interval 
estimation.

The code of \cite{GolightlyWilkinson:2005} was
  adapted to use the same half-Cauchy prior for the parameters as us and to
  employ an optimally-tuned single-block RWM Gaussian proposal
  so that this aspect would be exactly comparable with our RWM
  scheme. All MCMC algorithms except
  for that of \cite{GolightlyWilkinson:2005} were run for $110,000$
  iterations, from which the first $5,000$ iterations were discarded as
  burn-in. Output from simulations for
  the autoregulatory system was thinned by a factor of $10$ for storage. Since it
  mixed more slowly the
  algorithm of \cite{GolightlyWilkinson:2005} was run for $260,000$
  iterations with a burn-in of $10,000$ iterations; as with the
  LNA-based analysis of the autoregulatory system, output was thinned
  by a factor of $10$. 

In assessing accuracy of a method on a given model, throughout we
present results based on analysing 100 different data sets
simulated using the true jump process for 
a given model and set of parameter values. We present results in terms of estimating the log of the parameter since interest is
primarily in the order of magnitude of the reaction rates. The
posterior median is used as the point estimate for each parameter as this is
both invariant to monotonic transformations and is robust to heavy-tailed posterior distributions. 

\subsection{Comparison with ODE method and Komorowski et al.}
Our first comparison is with approximating the evolution as a
deterministic ODE, and with the LNA method of
Komorowski et al. and is based on the Lotka-Volterra
model. We simulated data with $\bmX_0=(40,140)$
 and with $\theta_1=0.01$, $\theta_2=0.6$ and $\theta_3=0.3$. Observations were made every second for 30 seconds, or until one of the
species became extinct.  We used Gamma(2,10) priors
for all rate parameters, which gave a reasonable prior mass across the range of values the rates take. We assumed that predators are observed
exactly for the LNA method and for that of
Komorowski et al., and the prey are unobserved.

The ODE method uses a log-likelihood (and hence log-posterior) which depends on the sum of squared errors between the solution of the ODE and 
the observations, and is equivalent to modelling the observations as
having additive Gaussian error; we place a
half-Cauchy prior on the error variance, $\pi(\sigma^2)\propto 1/\left(1+\sigma^4\right)$. Both LNA methods
have a similar computational cost, while the ODE approach has a smaller computational cost as the differential equations
for the variance in the LNA need not be solved.

\begin{table}
\caption{\label{Tab:1} Comparison of our approach (LNA), the approach of Komorowski et al. (KOM) and the ODE approach (ODE). We present
results in terms of the mean performance across 100 data sets. }
\begin{center}
\begin{tabular}{c|r|r|r|r}
&&$\theta_1$&$\theta_2$&$\theta_3$\\
&$\log_{10}\theta$&-2.000&-0.222&-0.523\\
\hline
Mean of&LNA&-2.001&-0.248&-0.541\\
posterior&KOM&-2.000&-0.263&-0.540\\
medians&ODE&-2.007&-0.308&-0.540\\
\hline
Mean abs.&LNA&0.043&0.056&0.050\\
error of&KOM&0.074&0.086&0.070\\
median&ODE&0.080&0.126&0.098\\
\hline
Mean&LNA&0.193&0.188&0.198\\
width of &KOM&0.138&0.153&0.156\\
95\% CI&ODE&0.141&0.158&0.194\\
\hline
Coverage&LNA&93&84&88\\
 of&KOM&51&54&55\\
95\% CI&ODE&43&29&30\\
\end{tabular}
\end{center}
\end{table}

Results are shown in Table \ref{Tab:1}. The new LNA approach is uniformly better at estimating each parameter in terms of both accuracy of point
estimates and coverage of credible intervals. The approach of Komorowski et al. is in turn more accurate than using an ODE approximation.
These results are for a small system, and the difference between the different approaches will be less if a larger system size is studied.  

\begin{figure}
\begin{center}
\centerline{\psfig{figure=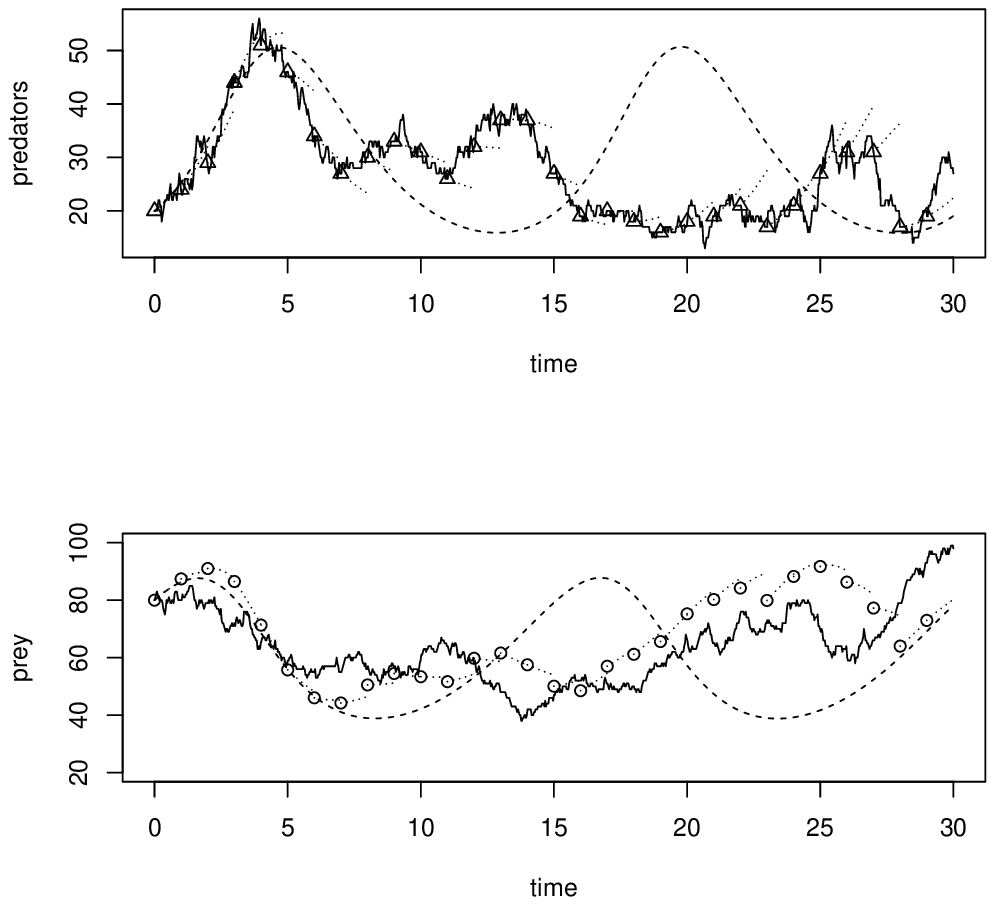,scale=.5,angle=270}}
\caption{\small{The true jump process (solid line) with observed values 
    marked as triangles and posterior mean values at observation times
    marked as circles. The dashed line shows the LNA 
solution for the deterministic process integrated forward from the
starting values, as used by Komorowski et al.; dotted lines show the LNA solution to the
deterministic process integrated from the observed or 
posterior mean value at each
observation time, as used in  our LNA method. The top plot corresponds to
predators (observed) the bottom plot to prey (unobserved).}
\label{fig.LNAKF.LV}
}
\vspace{-.5cm}
\end{center}
\end{figure}

To see why the new LNA approach is more accurate than either that of Komorowski et al. or that of using an ODE approximation, it is useful to see 
results from a single dataset. In Figure \ref{fig.LNAKF.LV} we show a simulated data set, together with the best fitting ODE solution. This solution gives
a poor fit to the data, and as the LNA of Komorowski is based on such ODE solutions, that approach gives a poor approximation to the likelihood. By comparison,
as our approach re-starts the LNA at each observation, we get a good approximation to the likelihood terms across the whole time-period of the
data.

\subsection{Comparison with SDE approach of Golightly and Wilkinson}
\label{sect.cmpGW}
We now compare our method with an approach based on an SDE approximation to the model \cite[]{GolightlyWilkinson:2005}. 
We compare on the auto-regulatory model (Example 2) with observations
every half a second for 25 seconds with the parameter values detailed in Table \ref{Tab:2}.
We considered models where all components of the system are observed without error, 
as the code that implements the approach of
\cite[]{GolightlyWilkinson:2005} assumes this observation model. For
both inference methods we assumed independent half-cauchy priors,  
$\pi(\theta_i)\propto 1/(1+(2\theta_i)^2)$ for $\theta_i>0$ for all $i$.

 The method  of \cite{GolightlyWilkinson:2005} involves imputing the path of the SDE at $m-1$ time-points in between each observation. 
We present results for $m=10$, which gave a good trade-off between accuracy and computational efficiency.

As well as comparing the accuracy of point and interval estimates for
the two methods, we also compare the computational efficiency. To do
this we calculated the integrated auto-correlation time for each parameter, and from this 
the Effective Sample Size (ESS). We summarise results in terms of the ESS divided by CPU time. 

\begin{sidewaystable}
\begin{center}
\caption{\label{Tab:2} Comparison of our LNA approach with that of the SDE-based method of Golightly and Wilkinson (2005). }
\begin{tabular}{c|r|r|r|r|r|r|r|r|r}
&&$\theta_1$&$\theta_2$&$\theta_3$&$\theta_4$&$\theta_5$&$\theta_6$&$\theta_7$&$\theta_8$\\
&$\log_{10}\theta$&-1.000&-0.155&-0.456&-0.699&-1.000&-0.046&-0.523&-1.000\\
\hline
Mean of&LNA&-0.921&-0.092&-0.466&-0.687&-0.856&-0.076&-0.523&-0.974\\
post. med.&GW05&-0.975&-0.135&-0.445&-0.643&-0.972&0.036&-0.506&-0.941\\
\hline
Mean abs. err.&LNA&0.147&0.139&0.093&0.090&0.197&0.183&0.082&0.107\\
of med.&GW05&0.107&0.107&0.097&0.103&0.100&0.120&0.086&0.118\\
\hline
Mean width&LNA&0.730&0.721&0.408&0.410&1.065&1.047&0.410&0.429\\
of 95\% CI&GW05&0.533&0.534&0.441&0.428&0.565&0.600&0.413&0.437\\
\hline
Coverage of&LNA&94&94&89&92&94&95&92&88\\
95\% CI&GW05&95&93&91&86&97&97&91&86\\
\hline
Mean &LNA&2.28&2.67&0.79&1.04&1.24&1.24&0.97&0.56\\
ESS/sec&GW05&0.18&0.18&0.60&1.10&0.09&0.08&0.68&0.82\\
\end{tabular}
\end{center}
\end{sidewaystable}


Results are given in Table \ref{Tab:2}.  The comparative performance
of the two methods varies with parameter. For the parameters
$\theta_3$, $\theta_4$, $\theta_7$ and $\theta_8$, both methods are
similar in terms of accuracy and of size and coverage of credible intervals. Coverage of credible intervals are consistent with their nominal size. Computational efficiency of the two methods is also very similar, with a mean ESS per second of 0.84 and 0.80 for the LNA and SDE methods respectively.

However inference on the remaining parameters differs considerably. The SDE method is uniformly more accurate, and has substantially smaller credible intervals in all cases. 
For both methods, coverage of credible intervals is close to $90\%$ for these parameters. The computational efficiency of the LNA is substantially higher for these parameters with the  ESS per second an order of magnitude greater.

The parameters for which the inferences of the two methods differ consist of two pairs of reaction rates: 
$\theta_1$ and $\theta_2$ are the rates of reversible reactions linked to the product  $\sDNA\cdot \sPP$, 
whereas $\theta_5$ and $\theta_6$  are the rates of reversible reactions linked to the dimerisation of the protein.  
As such we would expect difficulty in estimating each rate individually, as increasing both $\theta_1$ and $\theta_2$, say, 
together would lead to similar data. The larger credible intervals produced by the LNA are consistent with this. 
Investigating the performance of the SDE approximation shows that the
Euler discretisation becomes increasingly inaccurate as the parameter
vector increases. Changes in the state
  vector over a
  discretisation interval $\Delta t$ from a system with rate parameters $k\bmtheta$
  have exactly the same distribution as changes in the state vector
  over an interval $k\Delta t$ with rate parameters $\bmtheta$. For
  $k>1$ this decreases
  the accuracy of the estimated likelihood. In general therefore the Euler
  method will not estimate the posterior well for large rates.

To investigate this further we considered a second example, with all
rate parameters increased by a factor of 4. This gives insight into whether the method of \cite{GolightlyWilkinson:2005}  is producing appropriate credible intervals for these parameters, or whether the method is biased towards smaller parameter values, which happened to be consistent with the true parameter values used for the first simulation study.

Detailed results are given in Table 1 of Appendix D. We observe poor inference for $\theta_1$, $\theta_2$, $\theta_5$ and $\theta_6$ using the method of \cite{GolightlyWilkinson:2005}: 
accuracy is lower than using the LNA, and the credible intervals are too small, leading to coverage probabilities less than 0.5 in all cases. However, the method of \cite{GolightlyWilkinson:2005} 
does give more accurate inferences for the remaining parameters. This
is likely to be because the extra variability arising from the larger
rates means that the perturbations of the system from 
the ODE solution are no longer small. 

\subsection{Accuracy of the LNA method}
We further investigate the accuracy of the LNA, by repeating the analysis of the auto-regulatory example, but considering different observation models. 
We considered all components observed with error, and only three
components observed either exactly or with Gaussian error; errors
  for each species were independent with mean zero and variance $1$. We use the same priors as above.

Results are given in Appendix D and are comparable with
those in Table \ref{Tab:2}. The LNA appears to provide good estimates of the parameters. As we would expect, as
we observed fewer species, or observe with error, the uncertainty in our estimates increases. Perhaps most importantly, the coverage rates we obtain are close to 95\% in all cases, suggesting
the method is giving a good estimate of uncertainty.  
We obtain higher coverage rates with less informative data, possibly because any bias in the LNA has less effect when
we have higher posterior variance.

\section{Prediction of Flu Epidemics using Google Flu Trends Data}
\label{sec.GFT}
We now apply our method to predict flu case numbers based
on data from Google Flu Trends (GFT; http://www.google.org/flutrends).
GFT data are estimates of
the number of new cases of flu each week (per $100,000$ people) based on the popularity of
terms associated with flu 
in web
searches. \cite{Ginsbergetal:2009} showed that actual numbers of flu cases can be
acccurately predicted using such data, with the advantage of being
able to obtain estimates of the current number of cases, as opposed to
health-service data which 
  are
  typically 
  published with a delay of approximately one week  and are often
  incomplete.

Our analysis is motivated by \cite{JASA:2012}, who use a
one-compartment 
SEIR model (see Appendix A for details) to show that accurate
predictions of flu cases can be obtained from GFT data
using a state-space SEIR model, and in particular that such models are
substantially more accurate than simple AR models. For our
analysis we consider cases in the North and South
Islands of New Zealand. GFT data were
obtained for each island, for January 2008 to January 2012 inclusive,
and converted from proportions into counts.

Each year there is a flu epidemic, often with different flu
strains. The number of flu cases in
  New Zealand is typically at its yearly minimum around the start of
  February, and so we split our data
  into four separate `years'  from  February
  $yr$ to January of $yr+1$ for $yr\in\{2008,2009,2010,2011\}$.

We model the data using a two-compartment SEIR model. Our state
consists of the number of susceptibles, exposed, infected and
recovered in each of the north and south islands. We assume
a fixed population size for both islands, which results in a
6-dimensional state: $\{S_1,E_1,I_1,S_2,E_2,I_2\}$, where a subscript
of $1$ denotes North Island and a subscript of $2$ denotes South
Island. 

The reactions in our model are:
\begin{center}
\begin{tabular}{ll}
$R_1:$ $S_1+I_1\rightarrow E_1+I_1$
&
$R_2:$ $S_2+I_2\rightarrow E_2+ I_2$\\
$R_3:$ $E_1\rightarrow I_1$&
$R_4:$ $E_2\rightarrow I_2$\\
$R_5:$ $I_1\rightarrow 0$&
$R_6:$ $I_2\rightarrow 0$\\
$R_7:$ $S_1+I_2\rightarrow E_1+I_2$&
$R_8:$ $S_2+I_1 \rightarrow E_2+I_1$\\
\end{tabular}
\end{center}

\vspace{.5cm}
Further details of the model are provided in Appendix A.

Observations are $\bmy_t=(y_t^{(1)},y_t^{(2)})$, the number of flu cases, from the GFT data, in week $t$ in the North and South Island respectively. 
We model that these are realisations of random variables $Y_t^{(j)}$,
for $j=1,2$, where $Y_t^{(j)} \sim \mbox{N}(C I_j(t),\sigma^2_j)$. 
  A heuristic interpretation is that our SEIR models apply to the number of `communities' that are infected, and assume an equal rate of contact between each pair of communities. 
  If a community  is infected, then $C$ is the average number of flu cases that will result. 
Priors follow a similar form to those in \cite{JASA:2012};
 full details are
  given in Appendix E.

For each year's data, and for each $t=2,\ldots,51$ we use Steps
(1)-(3) in Section \ref{sect.approx.key.steps} within an adaptive RWM
algorithm similar to that in \cite{Sherlock/Fearnhead/Roberts:2010}
(see Appendix F) to estimate both the parameters and the current state of the model given
observations $\bmy_{1:t}$. For each sample from the RWM we then apply Step
(1) again to predict the number of cases in week $t+1$.   

As a benchmark to compare with, we also analysed the data using the
method of \cite{JASA:2012}. This approach uses sequential Monte
Carlo (SMC), and we shall refer to it as the SMC approach. It is based
 on fitting a single compartment SEIR model (henceforth 1CM) to data on the total number
 of flu cases across both islands. To deal with
 the intractability of the Markov jump model, the ODE approximation
 (\ref{eqn.deterministic.y}) is used and is itself approximated using an Euler
 scheme with a discretisation of the inter-observation interval (here
 one week). Gaussian noise is then added to the relative change in the
 number of infectives between observation times, 
leading to a very different Gaussian transition
 model to (\ref{eqn.transitionGauss}). The observation model is also based
 on the relative change in the number of infectives, with additive
 errors assumed to be Gaussian.
The SMC scheme approximates the 
 joint posterior for the parameters and the state at time $t$, given the data up to time
 $t$, by a set of weighted particles. Due to the choice of approximations
 of both the state dynamics and the observation, efficent methods \cite[]{Carvalho:2010,Pitt/Shephard:1999}
 for implementing the SMC algorithm can be used. See \cite{JASA:2012} for more details.
 
 One advantage of this approach is computational, as, unlike with MCMC, the algorithm does not need to be re-run from 
 scratch each time a new observation is received.  The potential
 disadvantage of the method is that it uses a cruder approximation to the underlying jump-Markov model.

We attempted to implement an equivalent SMC approach to
fit a two component SEIR model (henceforth 2CM) to the Google flu-trends data. However results from the SMC analysis of
this model, using $10^7$ particles, were substantially worse than for the 1CM. SMC methods are known to often perform poorly for models
with unknown parameters. The poor results for the 2CM are thus likely to be due to poor Monte Carlo performance for a model with 
10 unknown parameters. For further comparability with the method of
\cite{JASA:2012} we therefore also analysed data for the whole of New
Zealand using the LNA within a 1CM.


For each LNA analysis we ran an MCMC for 100,000 iterations,
using a burn-in of 20,000 for the 2CM and of 10,000
for the 1CM (then thinning both by a factor of
  $10$). 
For a week-ahead prediction at the height of the flu season (after 30
weeks of data) runs for the 1CM model took between 150 and
156 seconds on a single Intel Core i7 3770 CPU$@$3.40GHz, while runs
for the 2CM took between 1054 and 1135 seconds. 
Repeated runs of the MCMC produced the same estimates of accuracy to
at least two significant figures. The SMC analysis achieved a similar precision to the LNA when $10^6$
particles were used. Week-ahead
predictions from 30 weeks of data took between 67.7 and 68.1 seconds. 

\begin{table}
\caption{\label{Tab:GFT} Accuracy of one-week-ahead predictions for New Zealand
    as a whole for the one- and two-compartment models using the LNA
    and for the one-compartment model using SMC; average bias and mean absolute deviation (in cases per 100,000)
     and mean width (also in cases per 100,000, denoted MWCI) and coverage of the $95\%$ credibility interval.}

\begin{center}
\begin{tabular}{c|c|r|r|r|r}
Year &Method & Bias & MAD & MWCI & Cov. \\ \hline
2008 & LNA2CM & -2.01 & 6.03&21.8 & 84 \\
& LNA1CM& -1.11 & 5.96& 29.1& 84 \\
&SMC1CM & -3.07 & 6.95 & 185.7& 100\\ \hline
2009 & LNA2CM & 0.28 & 12.90& 36.6 & 84 \\
&LNA1CM & -0.27 & 14.72 & 40.51 & 86\\ 
& SMC1CM & -13.89 & 21.47& 211.3 & 100 \\ \hline
2010 & LNA2 & 0.08 & 6.42&19.4 & 82 \\
&LNA1CM & -0.37 & 6.29 & 21.6 & 84\\
& SMC1CM & -4.24 & 8.38& 113.8 & 100 \\ \hline
2011 & LNA2CM & -0.83 & 5.09 & 18.3 &84\\
& LNA1CM & -1.02 & 4.95 & 18.1&82\\
& SMC1CM & -1.50 & 5.82 & 92.2 &100\\ \hline
\end{tabular}
\end{center}
\end{table}

Summaries of the accuracy of both models using the LNA and of the
1CM using SMC in predicting
the total number of cases across both islands, and the coverage
of $95\%$ credible intervals for the predictions are given
in Table \ref{Tab:GFT}. 
Compared to the SMC method we see that, for all four years, the 
LNA (using either model) gives less biased estimates and smaller
forecast error, as
measured by the mean absolute error in predictions.
The credible intervals produced by the LNA methods are 
at least a factor of 5 smaller than for the approach
of \cite{JASA:2012}. However the coverage of the LNA's
credible intervals is between 80\% and 86\%. We believe the reason
for this is most likely due to our model assuming a constant variance for the observation error,
whereas the variance of this error appears to increase with the current size of the epidemic. We thus
under-estimate the uncertainty at the peak of the flu
epidemic. From the table, predictions for New Zealand as a whole from the 2CM are no
better than those from the 1CM, however the 2CM also provides
individual predictions for North and South Island.
These predictions, together with the true number of cases (as given by
the Google Flu Trends data), are shown in Figure
2 in Appendix D; summaries of the accuracy are provided in Table 3 of
the same appendix.

\section{Discussion}
\label{sect.discuss}

We have demonstrated how the LNA can be used to perform inference for reaction networks where all, or a subset, of components are observed. 
Observations can either be exact, or with additive Gaussian error.  Results suggest that using the LNA is more accurate than approximating the underlying model using an ODE.

The LNA is based upon first obtaining a deterministic approximation to the path of the state vector over time; and then modelling the error about this deterministic approximation. 
Key to the error in the LNA being small is that the deterministic approximation is accurate.  
We have shown that recalculating the deterministic solution between each pair of observations is more accurate than calculating a single deterministic solution as suggested by \cite{Komorowskietal:2009}.
Furthermore, across the examples we looked at the LNA gives reliable
inferences in almost all cases. The one exception (for a subset of
parameters in the results in Table 1 of Appendix C) corresponds to cases where the
noise in the model was high, leading to the peturbations of the system about the deterministic solution not being small.

We have also compared with a method based on approximating the underlying model using an SDE. The accuracy of the LNA and SDE-based approaches
are similar, with the relative performance of the two approaches varying depending on which reaction rates are being estimated.
The advantage of the LNA is one of simplicity -- as the LNA gives an analytic form for the approximation to the transition density of the model. 
In particular there is no need to choose a level of time-discretisation. 
Calculating the LNA involves solving a set of ODEs, but standard routines exist for appropriately choosing and adapting the step-size using in numerically solving the ODEs. 
By comparison SDE methods currently involve the user pre-specifying the level of time-discretisation. Choosing an appropriate level is difficult, partly because the required 
level needed to get an accurate approximation can depend on the parameter values, and these will change at each MCMC iteration. 

We have demonstrated the usefulness of the LNA for inference by making predictions for flu cases in New Zealand using Google Flu Trends data.  Whilst our prediction accuracy was higher than that of \cite{JASA:2012}, our assumption of a constant observation variance
leads to us under-estimating uncertainty in future observations at the peak of the epidemic. This assumption is currently needed for the tractability of our algorithm, but it should be possible to relax this assumption, for 
example using ideas from \cite{Rue/Martino/Chopin:2009} to allow for efficient inference under a range of observation models.  

 We considered fitting a two-compartment SEIR model to the New Zealand data, but the scalability of the LNA should mean it is possible to analyse SEIR 
models with even more compartments -- for example to jointly analyse
data from multiple cities in the US. For a reaction network with $n_r$ reactions
  and a state-space of size $n_s$, the LNA requires the numerical
  integration of $n_s^2$ ODEs, with
  $O(n_r)$ rate-related calculations at each time-point. For the
  two-compartment model, the state-space was twice the size of that of
  the one-compartment model, and the
  number of reactions more than doubled. Given the other computational
  overheads of the algorithm, this is consistent with the observed increase in
  CPU time by a factor of approximately 7, and, given the short 
  running time for the two-compartment model (less than 20 minutes),
  suggests that on-line
  week-ahead predictions should be feasible for models with three or
  four compartments. 

The approach of \cite{JASA:2012} uses a sequential Monte Carlo algorithm, which is computationally 
more convenient than MCMC. SMC inference for the single-compartment model of \cite{JASA:2012}
   was more than twice as fast as the single compartment
  LNA, but we could not implement an accurate SMC method for the two-component model. This seems
to be due to problems with using SMC to analyse models with moderately
large numbers of parameters.
Alternative sequential Monte Carlo approaches, such as those based on mode tracking \cite[e.g.][]{Vaswani:2008}, have recently been shown to be accurate for high-dimensional state processes,
and may offer a competitive alternative for analysing data such as that from Google Flu Trends.


\section{Supplementary Materials}
Web Appendices A, B, C, D, E and F, referenced respectively in
Sections~\ref{sect.reaction.networks}, \ref{sect.intro.thealgo},
\ref{sect.intro.thealgo}, \ref{sect.cmpGW}, \ref{sec.GFT}
and \ref{sec.GFT}, are available with
this paper at the Biometrics website on Wiley Online
Library.
The supplementary material also contains C code implementing the LNA for all of the models considered in
this article.

\backmatter

\section*{Acknowledgements} 
The authors are grateful to Dr. A. Golightly for
supplying C code for inference on the completely and exactly observed
autoregulatory example using the
methodology in \cite{GolightlyWilkinson:2005}, and to Dr. V. Dukic for
supplying the R code for prediction from the Google Flu Trends data
using the methodology in \cite{JASA:2012}. Paul Fearnhead was supported by EPSRC grant EP/G028745.

\vspace*{-8pt}

\bibliographystyle{biom}
\bibliography{sysbiol}

\label{lastpage}

\end{document}


\date{SomeDate}

\pagerange{\pageref{firstpage}--\pageref{lastpage}} 
\volume{TBA}
\pubyear{2014}
\artmonth{TBA}

\doi{blah}
\label{firstpage}

\maketitle

\appendix

\section{A: Epidemic Models}

\subsection{Example 3: An SEIR model}

The SEIR epidemic model \cite[]{Anderson/May:1991} describes the evolution of
an epidemic in a population. Each member of the population can be either
susceptible, exposed, infected or removed. We denote the number of people in each
of these states by $S$, $E$, $I$ and $\phi$ respectively. The reactions for this
model are
\begin{center}
\begin{tabular}{lll}
$R_1:$ $S+I\rightarrow E+I$
&
$R_2:$ $E\rightarrow I$\\
$R_3:$ $I\rightarrow \phi$ &
\end{tabular}
\end{center}
\vspace{.8cm}

If we assume the population is randomly mixing, so that each person is equally to likely to come into
contact with any others, then the rates for these reactions will be $(\theta_1 SI,\theta_2 E, \theta_3 I)$.

If we further assume a fixed population size then, as $S+E+I+\phi$ will remain constant, we
can define the state of the process as being just the number of people 
that are susceptible, exposed and infected, $(S,E,I)$. The resulting net effect matrix is
\[
 \bmA'=
\left[
\begin{array}{rrr}
-1&0&0 \\
1&-1&0\\
0&1&-1\\
\end{array}
\right]
\]

\subsection{Example 4: A two-component SEIR model}

Household models \cite[e.g.][Chapter 6]{Andersson/Britton:2000} are used to
model the spread of an infection which is transmitted at different
rates between different pairs of individuals. In particular,
infection spreads at a higher rate between individuals in the same
household than it does between individuals in different households. We
apply the same idea to the populations of two islands, say
$\mathsf{A_1}$ and $\mathsf{A_2}$, (rather than two
households). The four states for island $i$ correspond to the number
of susceptibles ($S_i$), exposed ($E_i$), infected ($I_i$) and removed
($R_i$) people who are resident on that island. The reactions for this
model are
\begin{center}
\begin{tabular}{lll}
$R_1:$ $S_1+I_1\rightarrow E_1+I_1$
&
$R_2:$ $S_2+I_2\rightarrow E_2+ I_2$\\
$R_3:$ $E_1\rightarrow I_1$&
$R_4:$ $E_2\rightarrow I_2$\\
$R_5:$ $I_1\rightarrow \phi_1$&
$R_6:$ $I_2\rightarrow \phi_2$\\
$R_7:$ $S_1+I_2\rightarrow E_1+I_2$&
$R_8:$ $S_2+I_1 \rightarrow E_2+I_1$
\end{tabular}
\end{center}

\vspace{.8cm}

Since the behaviour of
the inhabitants of each island $\mathsf{A_i}$ might well be different,
and different with respect to each other island we allow for four
different rates of transmission:
between individuals who
are resident in $\mathsf{A_1}$ ($\theta_1$), between individuals resident in
$\mathsf{A_2}$ ($\theta_2$), 
from individuals resident in $\mathsf{A_1}$ to those
resident in $\mathsf{A_2}$ ($\theta_6$) and from individuals resident in
$\mathsf{A_2}$ to those resident in $\mathsf{A_1}$ ($\theta_5$). 
The reaction rates are therefore
$(\theta_1S_1I_1,~\theta_2S_2I_2,~\theta_3E_1,~\theta_3 E_2, 
~\theta_4 I_1, ~\theta_4 I_2, ~\theta_5 S_1I_2, ~\theta_6 S_2I_1)$.

Assuming a fixed population size for each island, means that we can
reduce the dimension of the state-vector.
Taking this to be $(S_1,E_1,I_1,S_2,E_2,I_2)$, the
net-effect matrix is $\bmA$, where
\renewcommand{\arraystretch}{0.8}
\[
\bmA'=
\left[
\begin{array}{rrrrrrrr}
-1&0&0& 0&0&0&-1&0 \\
1&0&-1&0&0&0&1&0\\
0&0&1&0&-1&0&0&0\\
0&-1&0&0&0&0&0&-1\\
0&1&0&-1&0&0&0&1\\
0&0&0&1&0&-1&0&0
\end{array}
\right]
\]

\section{B: ODE solution for the LNA}

Define $\bmm\oft:=\Expect{\bmM\oft}$,
$\bmG\oft:=\Expect{\bmM\oft {\bmM\oft}^t}$ and 
$\bmPsi\oft:=\Var{\bmM\oft}=\bmG\oft-\bmm\oft \bmm{\oft}^t$.
Then by the linearity of expectation and the fact that
$d\bmM\oft=\bmF\oft \bmM\oft~dt + \bmS\oft ~d\bmW\oft$,
\begin{eqnarray*}
d\bmm\oft&=&\Expect{d\bmM\oft}=\Expect{\bmF\oft\bmM\oft~dt}=\bmF\oft\bmm\oft~dt.\\
d\bmG\oft&=&\Expect{\bmM\oftime{t+dt}\bmM\oftime{t+dt}^t-\bmM\oft\bmM{\oft}^t}\\
&=&\Expect{\bmM\oft d\bmM{\oft}^t+d\bmM{\oft}\bmM{\oft}^t+d\bmM\oft d\bmM{\oft}^t}
\\
&=& \Expect{\bmM\oft \bmM{\oft}^t\bmF{\oft}^t+ \bmF\oft \bmM\oft
  \bmM{\oft}^t}dt\\
&& + \Expect{\bmS\oft\bmS{\oft}^t}dt\\
&=& \bmG\oft\bmF{\oft}^t+\bmF{\oft}\bmG\oft+\bmS\oft\bmS{\oft}^t.\\
d\bmPsi\oft&=&d\bmG\oft - \bmm\oft~ d\bmm{\oft}^t-d\bmm\oft~
\bmm{\oft}^t\\
&=&\bmPsi\oft\bmF{\oft}^t+\bmF\oft\bmPsi\oft+\bmS\oft\bmS{\oft}^t, 
\end{eqnarray*}
as required.

\section{C: Estimation of transition probabilities}
The SDE and the LNA can be viewed as nested approximations to the
evolution of the
reaction network. The accuracy of the transition densities arising
from these two approximations is now
investigated for the autoregulatory model (Example 2). Both the LNA and the SDE arise from a continuous
approximation to a discrete process, with the LNA also assuming that
 stochastic variations are small compared to the magnitude of the
 process itself, and we might therefore expect both approximations to improve as the
 system size increases. Three different system
sizes are therefore compared with initial conditions
$\bmx\oftime{0}=(5\Omega,8\Omega,8\Omega,8\Omega,5\Omega)$ for $\Omega\in \{1,10,100\}$.

The rate parameter vector was  set to
$\bmtheta=(0.1/\Omega,0.7,0.35,0.2,0.1/\Omega, 0.9,0.3,0.1)$
so as to keep the behaviour of the system (and of the drift ODE)
consistent with the system examined in \cite{GolightlyWilkinson:2005},
which uses the  above initial state vector and rate vector with 
$\Omega=1$. The true system was simulated forward $10000$ times and the distribution of species was stored after  
$t_1=0.1$ time unit, $t_2=0.5$ time unit and $t_3=2.5$ time
units. The SDE was integrated forward  $10000$ times using an
Euler-Maruyama time step of $0.001$ time units. Finally the
parameters for the multivariate Gaussian transition density for $t_1$, $t_2$ and
$t_3$ were estimated by integrating forward the drift and variance ODEs.
\begin{figure}
\begin{center}
\centerline{\psfig{figure=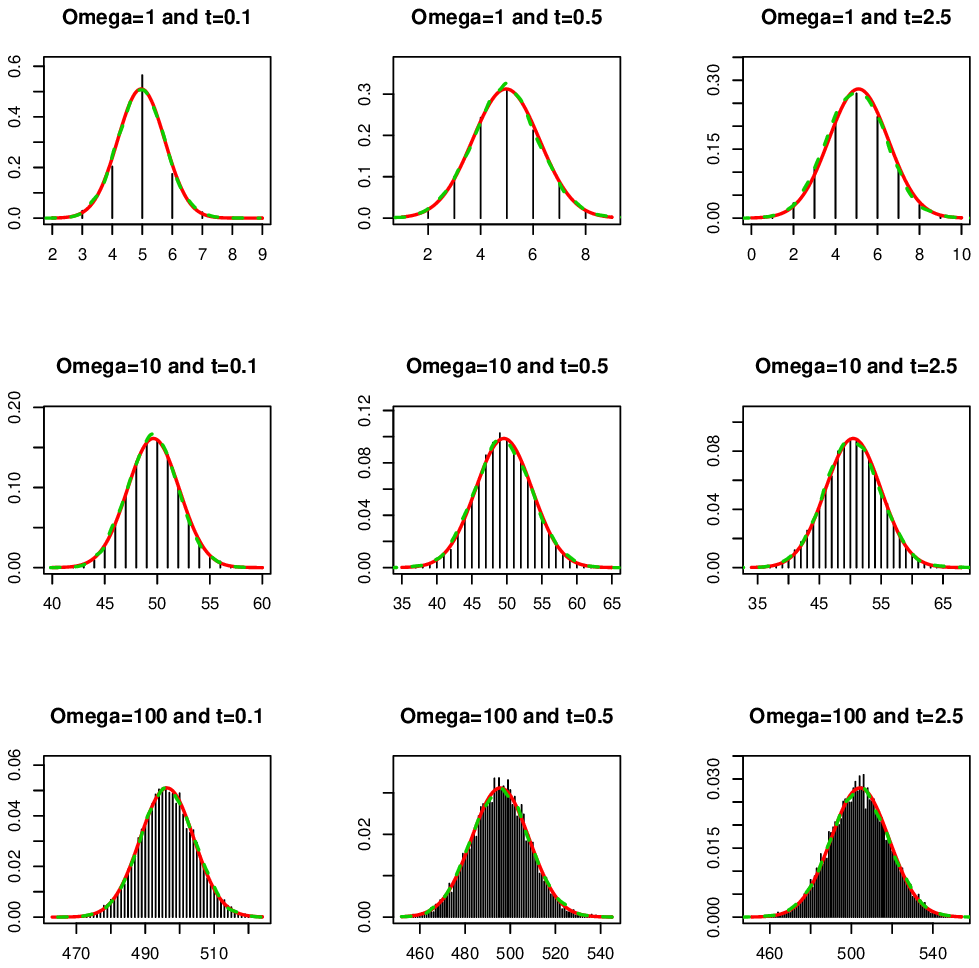,scale=.5,angle=0}}
\caption{\small{Transition probabilities for DNA (species $1$)
      under rate vector $\bmtheta=(0.1/\Omega,0.7,0.35,0.2,0.1/\Omega,
      0.9,0.3,0.1)$ from an initial state of $(5\Omega,8\Omega,8\Omega,8\Omega,5\Omega)$ for $\Omega\in \{1,10,100\}$  estimated from the
    LNA (solid red line) and from $10000$ simulations each for the true process (solid
    black bars) and
     the CLA (dashed green line) with an Euler timestep of $\Delta t=0.001$. System
    sizes are $\Omega=1$ (top row), $\Omega=10$ (middle row) and
    $\Omega=100$ (bottom row), and times
    are $t=0.1$
    (left column) $t=0.5$ (middle column) and $t=2.5$ (right
    column).  For output from the exact simulation 
 vertical bars are used to represent
    the relative probability of each outcome; kernel density estimates
    are used for the CLA, whilst for the LNA the plot simply shows the
    Gaussian density with the mean and variance estimated using the
    LNA. The $y$ axis represents probability mass for the true process
    and probability density for the two continuous
    approximations. However $x$ copies of a species simulated from the true process
    corresponds to between $x-1/2$ and $x+1/2$ copies using the
    continuous approximation. Since this interval has width $1$ the
    probability mass and probability density plots are directly comparable.}
\label{fig.trans.prob.sim}
}
\vspace{-.5cm}
\end{center}
\end{figure}

Comparison of the transition parameters are shown in Figure \ref{fig.trans.prob.sim}. 
The error induced by representing the true system with a continuous approximation can be seen for the small and medium system sizes. 
However, throughout both the LNA and SDE models give, by eye, a reasonable approximation to the true transition density and
 there appears to be little difference in the accuracy of the two
 approximations. The main difference is that the LNA always gives a
 Gaussian approximation, whereas the SDE transition density can be
 non-Gaussian. Graphs for the other species showed the same
   pattern and comparisons of bivariate distributions all showed good agreement. Qualitatively similar results have been observed for a variety of models and parameter values; see \cite{Giagos:2011}.

\section{D: additional tables from the simulation study
  for the autoregulatory model}

\begin{sidewaystable}
\begin{center}
\caption{\label{Tab:3} Comparison of our LNA approach with that of the SDE-based method of Golightly and Wilkinson (2005). 
The true reaction rates are all four times larger than for Table 2 in
the main text}
\begin{tabular}{c|r|r|r|r|r|r|r|r|r}
&&$\theta_1$&$\theta_2$&$\theta_3$&$\theta_4$&$\theta_5$&$\theta_6$&$\theta_7$&$\theta_8$\\
&$\log_{10}\theta$&-0.398&0.447&0.146&-0.097&-0.398&0.556&0.079&-0.398\\
\hline
Mean of&LNA&-0.592& 0.229&-0.099&-0.324& -0.596& 0.324&-0.164&-0.638\\
post. med.&GW05&-0.674& 0.160&0.180& -0.037&-0.804& 0.219&0.090&-0.345\\
\hline
Mean abs. err.&LNA& 0.216&0.235&0.257&0.233&0.205&0.236&0.252&0.246\\
of med.&GW05&0.276&0.287&0.110&0.086&0.407&0.337&0.102&0.091\\
\hline
Mean width&LNA&1.258&1.263&0.559& 0.669&1.359&1.356&0.564&0.716\\
of 95\% CI&GW05&0.500&0.502&0.490&0.350&0.460&0.500&0.458&0.360\\
\hline
Coverage&LNA&100&100&59&79&99&99&59&75\\
 of 95\% CI&GW05&40&40&90&90&0&5&93&91\\
\end{tabular}
\end{center}
\end{sidewaystable}

\begin{sidewaystable}
\caption{\label{Tab:4} Accuracy of our LNA approach on the auto-regulatory example. 
Results are for all components observed with Gaussian error (4GE) and three component observed either exactly (3NE) or with Gaussian error (3GE) }
\begin{center}
\begin{tabular}{c|r|r|r|r|r|r|r|r|r}
&&$\theta_1$&$\theta_2$&$\theta_3$&$\theta_4$&$\theta_5$&$\theta_6$&$\theta_7$&$\theta_8$\\
&$\log_{10}\theta$&-1.000&-0.155&-0.456&-0.699&-1.000&-0.046&-0.523&-1.000\\
\hline
Mean of&4GE&-0.860&-0.035 &-0.439& -0.612& -0.842& 0.086&-0.497&-0.905\\
posterior&3NE&-0.737&-0.147&-0.320&-0.542&-0.843& 0.088&-0.539&-0.821\\
medians&3GE&-0.623&-0.117&-0.279& -0.555& -0.833& 0.093&-0.541&-0.833\\
\hline
Mean abs.&4GE&0.211& 0.202&0.153 &0.195 &0.207&0.192 &0.143& 0.212\\
error of&3NE&0.289&0.157&0.170&0.223&0.206&0.190&0.083&0.229\\
median&3GE&0.395&0.142&0.205&0.234&0.217& 0.201&0.158& 0.239\\
\hline
Mean&4GE&1.044&1.043& 0.880&1.079&1.215&1.202&0.808&1.180\\
width of&3NE&1.340&1.187&0.829&1.154&1.217&1.202&0.422&1.128\\
95\% CI&3GE&1.885&1.793&1.242& 1.264& 1.325& 1.314& 0.893& 1.221\\
\hline
Coverage&4GE&93&93&97&95&96&98&96&91\\
 of&3NE&94&100&97&90&96&98&94&92\\
95\% CI&3GE&100&100&96&96&97&99&97&95\\
\end{tabular}
\end{center}
\end{sidewaystable}

\section{E: priors and additional results for the GFT analysis}
The priors for our rate parameters use the independent truncated Gaussian
distributions employed in \cite{JASA:2012}
 with the same (Gaussian) mean as that paper but with the standard deviation
doubled to represent our uncertainty in the values of these parameters
for communities as opposed to individuals. 
All four transmission rates ($\theta_1,\theta_2,\theta_5,\theta_6$)
have $\theta\sim N(1.5,1)\ind{\theta>0}$. The rates for changing from
exposed to infected is $\theta_3\sim N(2,1)\ind{\theta_3>0}$ and the recovery
rate is $\theta_4\sim N(1,1)\ind{\theta_4>0}$.

For island specific variables let $i=1,2$ correspond to N and S Islands
repsectively. Let $N_i$ be the population of island $i$ which is assumed fixed
at $(3366100,1038400)$ over the period examined.

We follow a similar format to \cite{JASA:2012} for all other parameters and
quantities, and assign Gamma
priors with a shape parameter of $1.1$; these are relatively vague yet
 tail off to zero as the parameter or other quantity
approaches zero.

We have the prior $C\sim Gam(1.1,1.1/10)$.
Then the initial number of susceptibles is independent for each
island conditional on $C$:
\[S_i(0)|C\sim Gam(1.1,1.1/\mu_i)~\mbox{ where } \mu_i=N_i/(2C).\]
Subject to the constraint
\[
\ind{S_1(0)C\le N_1}\ind{S_2(0)C\le N_2}.
\]

The variance of the observations is independent for each island: 
\[\sigma^2_i \sim
Gam(1.1,1.1/V_i), \mbox{ where }V_i=\left(\frac{5N_i}{10^5
    }\right)^2.\]
The noise in the data (in cases per $10^5$) for the year preceding our analysis
appears to have a standard deviation of around $5$ (we simply need an
estimate of the order of magnitude for a sensible prior). The above transforms this to the variance on a pseudo observation of
number of communities infected. 

The initial state for island $i$ is Gaussian (as required by our
methodology) with
\[
\Expect{X_i(0)}=\left[S_i(0),\frac{a_iN_i}{2\times 10^5 C},\frac{a_iN_i}{10^5 C}\right],
\]
where $a_i$ is the observed number of cases per $10^5$ in the last
week in January (i.e. the week prior to the start of the data
file). We use the fact that the exposure rate is believed to be about
twice the infected rate. The variance matrix is diagonal with the
following entries
\[
\mbox{diag}\left[0,10\times\left(\frac{N_i}{10^5 C}\right)^2,10\times\left(\frac{N_i}{10^5 C}\right)^2\right].
\]

\begin{figure}
\begin{center}

\centerline{\psfig{figure=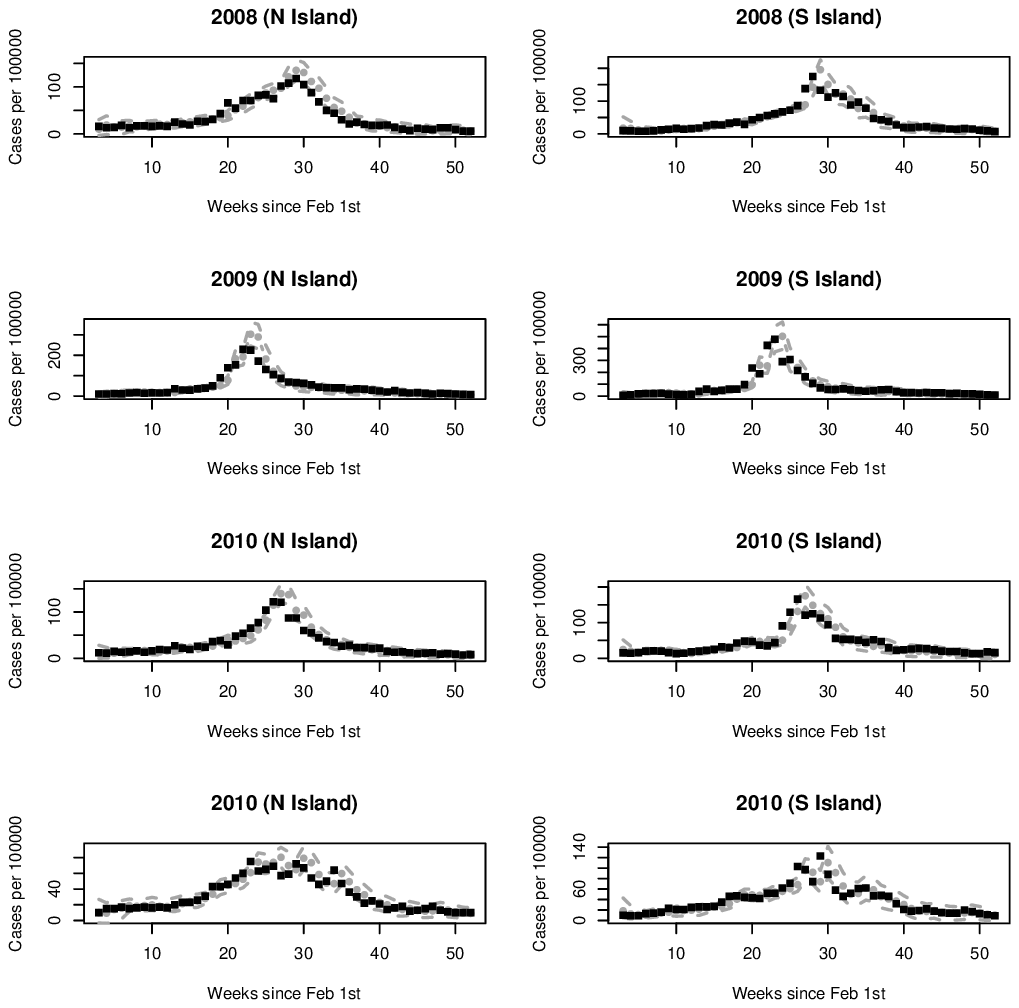,scale=.5,angle=0}}
\caption{\small{Observed influenza cases per 100000 from the GFT data
    for North and South Islands for 2008, 2009, 2010, and 2011 (black
    squares). Overlaid in grey are posterior median predictions using the
    LNA with a two-compartment SEIR model (circles and the central line) and
    upper and lower $95\%$ credible bounds (dashed lines). }
\label{Fig:GFT}
}
\vspace{-.5cm}
\end{center}
\end{figure}

\begin{table}
\caption{\label{Tab:GFTb} Accuracy of one-week-ahead predictions for New Zealand
    North and South Islands for the two-compartment model; average bias and mean absolute deviation (in cases per 100,000)
     and mean width (also in cases per 100,000, denoted MWCI) and coverage of the $95\%$ credibility interval.}
\begin{center}
\begin{tabular}{c|c|r|r|r|r}
Year &Island & Bias & MAD & MWCI & Cov. \\ \hline
2008 & N & -2.45 & 6.92&25.6 & 84 \\
& S& -0.44 &7.59 & 28.3 & 88 \\ \hline
2009 & N & 0.16 & 12.56& 38.0 & 90 \\
&S & 0.82 & 19.32 & 63.6 & 84\\ \hline
2010 & N & -0.35 & 6.44&21.6 & 78 \\
&S & 1.50 & 9.14 & 29.7 & 80\\ \hline
2011 & N & -1.23 & 5.40 & 20.7 &90\\
& S & 0.51 & 7.39 & 28.8&88\\ \hline
\end{tabular}
\end{center}
\end{table}

\section{F: the adaptive RWM algorithm used to fit the SEIR models}
The adaptive RWM algorithm updates the variance and overall scaling of
the random walk Metropolis proposal distribution according to the history of the Markov
chain. It is very similar to the algorithm in
\cite{Sherlock/Fearnhead/Roberts:2010}, except that adaptations to the
scaling parameter are
relative rather than absolute.

Let $\theta$ denote the current (p-dimensional) 
parameter vector and $\theta^*$ the
proposed new parameter vector. 
To ensure that any variance matrix calculated from the history of the
chain is non-singular, the algorithm proceeds with a fixed kernel ($\theta^*\sim
N_p(\theta,\Sigma_0)$) 
 until the number of
proposals that have been accepted is at least $2p^2$. 

At each subsequent iteration, $i$, the proposal is sampled from a mixture
distribution
\[
\theta^* \sim 
\left\{
\begin{array}{lll}
N_p(\theta,\Sigma_0)&w.p.&\beta\\
N_p(\theta,\lambda_i^2\Sigma_i)&w.p.&1-\beta
\end{array}
\right.,
\] 
for some $\beta\in (0,1)$ (we chose $\beta=0.05$). Here $\Sigma_i$ is
the variance of the MCMC sample from $\theta$ up to and including
iteration $i-1$. The adaptive scaling parameter, $\lambda_i$ is
initialised to $\lambda_0=1/\sqrt{p}$ and is altered each time 
 the adaptive proposal distribution
$N_p(\theta,\lambda_i^2\Sigma_i)$ is used, at which point the update
depends upon whether or not the proposal, $\theta^*$, was accepted:
\[
\lambda_{i+1}=
\left\{
\begin{array}{ll}
\lambda_i\left(1+2.3\delta/\sqrt{n}\right)
&\mbox{if $\theta^*$ was accepted}\\
\lambda_i\left(1-1\delta/\sqrt{n}\right)
&\mbox{if $\theta^*$ was rejected}\\
\end{array}
\right.
\]
Here $n\le i-2p^2$ is the number of adaptive proposals that have been
used so far, and $\delta$ is a user parameter which we set to
$0.1$. The algorithm targets an acceptance rate of $1/3.3\approx 0.3$.
\bibliographystyle{biom}
\bibliography{sysbiol}

\label{lastpage}